\documentclass[prl,showpacs,showkeys,nofootinbib,twocolumn,floatfix]{revtex4}

\usepackage{ulem}
\usepackage{graphicx}  %
\usepackage{amsmath}
\usepackage{bm}  %

\newcommand{\bea}{\begin{eqnarray}}
\newcommand{\eea}{\end{eqnarray}}
\newcommand{\beq}{\begin{equation}}
\newcommand{\eeq}{\end{equation}}
\newcommand{\bqa}{\begin{eqnarray}}
\newcommand{\eqa}{\end{eqnarray}}

\def\mqo2{{\!\!\!}}

\begin{document}

\title{
Atom Loss Resonances \\
in a Bose-Einstein Condensate}

\author{Christian Langmack}
\affiliation{Department of Physics,
         The Ohio State University, Columbus, OH\ 43210, USA}

\author{D.~Hudson Smith}
\affiliation{Department of Physics,
         The Ohio State University, Columbus, OH\ 43210, USA}

\author{Eric Braaten}
\affiliation{Department of Physics,
         The Ohio State University, Columbus, OH\ 43210, USA}

\date{\today}

\begin{abstract}
Atom loss resonances in ultracold trapped atoms have been observed 
at scattering lengths near {\it atom-dimer resonances}, 
at which Efimov trimers cross the atom-dimer threshold,
and near {\it two-dimer resonances}, 
at which universal tetramers cross the dimer-dimer threshold.
We propose a new mechanism for these loss resonances in a Bose-Einstein
condensate of atoms.
As the scattering length is ramped to the large final value 
at which the atom loss rate is measured, 
the time-dependent scattering length generates a small condensate of shallow dimers 
coherently from the atom condensate.
The coexisting atom and dimer condensates can be described 
by a low-energy effective field theory
with universal coefficients that are determined by 
matching exact results from few-body physics.
The classical field equations for the atom and dimer condensates
predict narrow enhancements in the atom loss rate 
near atom-dimer resonances and near two-dimer resonances
due to inelastic dimer collisions.
\end{abstract}

\smallskip
\maketitle

Nonrelativistic particles whose scattering lengths are large 
compared to the range of their interactions exhibit universal 
low-energy behavior \cite{Braaten:2004rn}. 
The universal few-body phenomena can include 
a spectrum of loosely-bound molecules as well as 
reaction rates of the particles and molecules.
In some cases, including identical bosons, the universal behavior 
is governed by discrete scale invariance.
The universal molecules then include sequences of 
3-particle clusters (Efimov  trimers) \cite{Efimov70,Efimov73}, 
4-particle clusters (universal tetramers)
\cite{Platter:2004qn,Hammer:2006ct,vSIG:0810},
and clusters with even more particles \cite{vonStecher:1106} .
The universal reaction rates exhibit intricate resonance and 
interference features \cite{Braaten:2004rn}.

The technology for trapping atoms and cooling them to extremely 
low temperatures has made the universal low-energy region
experimentally accessible.  The use of Feshbach resonances 
to control the scattering length $a$ experimentally makes ultracold atoms 
an ideal laboratory for universal physics.  
One particularly dramatic probe 
of universality is the loss rate of atoms from a trapping potential. 
Resonance and interference effects in few-body reaction rates 
can produce local maxima and minima in the loss rate as a function of $a$.
The most dramatic signature for a universal $N$-atom cluster with $N \ge 3$
is the resonant enhancement of the $N$-atom inelastic collision rate
at a negative $a$ where the cluster crosses the 
$N$-atom threshold and becomes unbound.
We refer to such a scattering length as an {\it $N$-atom resonance}.
The first observations of an Efimov trimer \cite{Grimm:0512} and 
a universal tetramer \cite{Grimm:0903} and the first evidence for
a universal 5-atom cluster \cite{Grimm:1205} were all obtained using
a thermal gas of $^{133}$Cs atoms by tuning $a$
to 3-atom, 4-atom, and 5-atom resonances, respectively.
An Efimov trimer has also been observed as an enhancement in the loss 
rate in a mixture of $^{133}$Cs atoms and dimers \cite{Grimm:0807}
at a positive scattering length $a_*$ where an Efimov trimer 
crosses the atom-dimer threshold and becomes unbound.  
We refer to $a_*$ as an {\it atom-dimer resonance}.  
Another dramatic loss feature at positive $a$
is an interference minimum in the 
3-atom recombination rate into the shallow dimer,
which was also first observed in a thermal gas 
of $^{133}$Cs atoms \cite{Grimm:0512}.
Many of these loss features have been subsequently observed 
in ultracold trapped atoms of other elements \cite{Ferlaino:1108}.

There are a few loss features in ultracold atoms 
that have not yet been related to universal few-body reaction rates.
They all appear in systems that were believed to consist of atoms
only and no dimers.
Narrow enhancements of the loss rate near atom-dimer resonances
have been observed in both a Bose-Einstein condensate (BEC) and a thermal gas
of $^{39}$K atoms \cite{Zaccanti:0904} and in both a BEC 
and a thermal gas of $^7$Li atoms \cite{Hulet:0911,Khaykovich:1201}.
In a BEC of $^7$Li atoms, narrow enhancements of the loss rate 
have also been observed at positive values of $a$
near {\it two-dimer resonances} \cite{Hulet:0911},
at which universal tetramers cross the dimer-dimer threshold and become unbound.
No mechanism has been proposed for a narrow loss feature
at a two-dimer resonance in a system consisting of atoms only and no dimers.
One proposed mechanism for a narrow loss feature near an atom-dimer 
resonance $a_*$ in such a system is the {\it avalanche mechanism}, 
in which the 3-body recombination rate 
into the shallow dimer is amplified by secondary elastic collisions 
of the outgoing dimer \cite{Zaccanti:0904}.
It was recently shown that the avalanche mechanism cannot produce a 
narrow loss feature near $a_*$ \cite{LSB:1205}.
This is a consequence of the universal energy dependence of 
the atom-dimer cross section.  Instead of having a narrow peak at $a_*$,
the elastic cross section for the energetic dimer from the recombination 
event has a broad maximum near $4.3\, a_*$.
Detailed Monte Carlo simulations of the avalanche process demonstrate
that it does not produce any narrow loss features \cite{LSB:1209}.

In this Letter, we propose a new mechanism for narrow loss
features near atom-dimer resonances and near two-dimer resonances
in a Bose-Einstein condensate of atoms.
The mechanism is motivated by the phenomenon of {\it atom-molecule coherence},
which involves the coherent transfer of atom pairs
between an atom condensate and a coexisting dimer condensate.
A small condensate of shallow dimers can be produced coherently by the
time-dependent scattering length as it is ramped to the large final value 
where the atom loss rate is measured.
The loss features then arise from the resonant enhancement of 
inelastic collisions involving dimers from the dimer condensate.

The phenomenon of atom-molecule coherence was discovered by 
Donley {\it et al.}\ in 2002 using a BEC
of $^{85}$Rb atoms \cite{Weinman:0204}.
In these experiments, a pulse in the magnetic field 
brought the atoms very close to a Feshbach resonance.
The atoms were allowed to evolve at a large constant 
scattering length for a variable holding time,
and then a second pulse took the atoms close 
to the Feshbach resonance again.
Subsequent measurements of the number of atoms
revealed three distinct components:
a ``remnant'' BEC, a ``burst'' of relatively energetic atoms,
and  ``missing'' atoms that were not detected.
The numbers of remnant, burst, and missing atoms all varied 
sinusoidally with the holding time at the frequency  
associated with the dimer binding energy.
That sinusoidal dependence 
can be explained by a coexisting condensate of shallow dimers
that was created by the first pulse \cite{KH:0204,GKB:0209,DS:0312}.

The behavior of atoms and dimers with sufficiently
small kinetic and potential energies 
can be described by a quantum field theory with 
independent fields for the atoms and dimers.
Atom and dimer condensates are described by classical fields
$\psi(\bm{r},t)$ and $d(\bm{r},t)$ that are the expectation values of the
quantum fields.  The quantum field equations can be formulated 
as coupled integro-differential equations for $\psi$, $d$, 
and an infinite hierarchy of correlation functions for 
quantum fluctuations.
A typical experiment on the atom loss rate in a Bose-Einstein condensate
begins with a stable BEC of atoms 
with a small positive scattering length.  
This system is described by a static atom condensate $\psi(\bm{r})$,
with $d(\bm{r})=0$ and zero correlation functions.
A ramp in the magnetic field produces a time-dependent scattering length
with a large final value $a$.
During the ramp, the system 
is described by time-dependent condensates $\psi(\bm{r},t)$
and $d(\bm{r},t)$ and nonzero correlation functions.  
At the end of the ramp, the dimer condensate $d(\bm{r},0)$ will be nonzero,
although it could be very small if the ramp is nearly adiabatic. 
It could presumably be calculated using the methods developed 
in Refs.~\cite{KH:0204,GKB:0209,DS:0312} to
describe atom-molecule coherence.
We will not attempt to calculate $d(\bm{r},0)$, but simply take
the initial fraction $f_D$ of the atoms that are 
bound into dimers in the dimer condensate to be an unknown initial
condition.  We will ask whether observable loss features 
can be produced by the dimer condensate for a plausibly small fraction $f_D$.

During the subsequent holding time, 
there are transient effects in the atom and dimer condensates 
and in the correlation functions that will die away.
We will assume that after they have died away,
the system can be described by coexisting atom and dimer condensates
only and that the correlation functions can be neglected. 
Atom-molecule coherence will produce oscillations in $\psi(\bm{r},t)$
and $d(\bm{r},t)$ at the frequency 
$\hbar/(2 \pi ma ^2)$ associated with the dimer binding energy. 
The time-averaged condensates also decrease with time due to loss processes.
The coexisting condensates can be described by a 
low-energy effective field theory for atoms and dimers 
whose kinetic and potential energies are small compared 
to the dimer binding energy.
The interaction terms in the classical Hamiltonian density include
\begin{eqnarray}
{\cal H}_{\rm int} &=&
\nu_2 d^* d + \frac{\hbar^2 \lambda_2}{4m} (\psi^* \psi)^2
+ \frac{\hbar^2 h_3}{m} (d^* d) (\psi^* \psi)
\nonumber
\\
&& + \frac{\hbar^2 f_4}{4m} (d^* d)^2 
+ \frac{\hbar^2 \lambda_3}{36m} (\psi^* \psi)^3 
 + \ldots .
\label{eq:H-eff}
\end{eqnarray}
The coefficients of the interaction terms are universal functions of $a$ and the complex
Efimov parameter $\kappa_* \exp(i \eta_*/s_0)$,
which can be interpreted as the binding wavenumber of an 
Efimov trimer in the unitary limit $a = \pm \infty$ \cite{Braaten:2004rn}.
The small positive parameter $\eta_*$ takes into account 
tightly-bound diatomic molecules (deep dimers),
which provide decay channels for Efimov trimers.
The coefficients of the interaction terms in Eq.~(\ref{eq:H-eff})
are constrained by discrete scale invariance to be
log-periodic functions of $\kappa_*$, with discrete scaling factor
$e^{\pi/s_0}$, where $s_0 \approx 1.00624$ is a universal constant.

The coefficients in Eq.~(\ref{eq:H-eff}) 
can be determined by demanding that few-body results 
in the fundamental theory be reproduced by the effective field theory.
The coefficient of $d^* d$ is determined 
by matching the rest energy of a shallow dimer:
$\nu_2 = - \hbar^2/m a^2$.
The coefficient of the terms that are 4$^{\rm th}$ order in the fields 
can be determined by matching elastic 
scattering amplitudes at threshold:
$\lambda_2 = 8 \pi a$, $h_3 = 3 \pi a_{AD}$, and $f_4 = 4 \pi a_D$,
where  $a_{AD}$ and $a_D$ are the atom-dimer and dimer scattering lengths.
The atom-dimer scattering length $a_{AD}$ is $a$ multiplied by a simple
log-periodic function of $a \kappa_*$  \cite{Braaten:2004rn}.
In the limit $\eta_*=0$, that function is real 
and it diverges at the atom-dimer resonance $a_{*} = 0.07076/\kappa_*$.
The dimer scattering length $a_D$ is $a$ multiplied by a
log-periodic function of $a \kappa_*$ that is complex 
even if $\eta_*=0$.  For $\eta_* = 0$, Deltuva has calculated
$a_D$ with several digits of accuracy \cite{Deltuva:1107}.
Its imaginary part has narrow resonant peaks at the two-dimer resonances 
$a_{1*} \approx 2.196\, a_*$ and $a_{2*} \approx 6.785\, a_*$.
These two-dimer resonances were first calculated in Ref.~\cite{vSIG:0810}.
For $\eta_*>0$, $a_D$ can be obtained by constructing an analytic fit to 
Deltuva's results for $a_D/a$ as a function of $a \kappa_*$ and then carrying out the
analytic continuation $\kappa_* \to \kappa_* \exp(i \eta_*/s_0)$.
The coefficient of $(\psi^* \psi)^3$ in Eq.~(\ref{eq:H-eff}) 
could be determined by matching the 
elastic 3-atom scattering amplitude in the low-energy limit.
By the optical theorem, its imaginary part is proportional to 
the 3-atom recombination rate coefficient, 
${\rm Im} \, \lambda_3 = - (3 m/\hbar) \alpha$,
which can be separated into contributions from recombination 
into the shallow dimer and into deep dimers:
$\alpha = \alpha_{\rm shallow} + \alpha_{\rm deep}$.
They both have the form $\hbar a^4/m$ multiplied by
log-periodic functions of $a$ \cite{Braaten:2004rn}.
In the limit $\eta_*=0$, $\alpha_{\rm deep}=0$
and $\alpha_{\rm shallow}$ has an interference zero at 
$a_{+} = 0.31649/\kappa_*$.

The time dependence of $\psi(\bm{r},t)$ and $d(\bm{r},t)$
is determined by the classical field equations associated with
the interaction Hamiltonian density ${\cal H}_{\rm int}$ 
in Eq.~(\ref{eq:H-eff}).
The corresponding number densities  are $n_A  = \psi^* \psi$ and $n_D = d^* d$.
The rates at which the numbers $N_A  = \int d^3r \, n_A$
and $N_D = \int d^3r \, n_D$  change is determined by 
the anti-hermitian part of ${\cal H}_{\rm int}$:
\begin{eqnarray}
\frac{dN_A}{dt} &=& 
\frac{6 \pi \hbar\, {\rm Im} a_{AD}}{m} \!\!\int \!\! d^3r \, n_D n_A
- \frac{\alpha}{2} \!\!\int \!\!d^3r \, n_A^3 ,
\label{eq:dN/dt}
\\
\frac{dN_D}{dt} &=& 
\frac{6 \pi \hbar\, {\rm Im} a_{AD}}{m} \!\!\int \!\! d^3r \, n_D n_A
+ \frac{4 \pi \hbar\, {\rm Im} a_D}{m} \!\!\int \!\!d^3r \, n_D^2  .
\nonumber
\end{eqnarray}
Since unitarity requires the imaginary parts of $a_{AD}$ and $a_D$ to be negative,
these equations imply that $N_A$ 
and $N_D$ both decrease monotonically with time.
This excludes atom-molecule coherence,
which involves coherent oscillations between $N_A$ and $N_D$
with angular frequency $E_D/\hbar$. 
The appropriate interpretation is that the high-frequency
variations in the condensates associated with atom-molecule coherence
are not resolved within our low-energy effective field theory. 
It can at best describe number densities $n_A(\bm{r},t)$ 
and $n_D(\bm{r},t)$ that are averaged over many periods
of the atom-dimer oscillations.

To predict the loss rate of atoms during the holding time,
we need initial conditions $\psi(\bm{r},0)$ and $d(\bm{r},0)$
that are robust approximate solutions of the classical field equations
associated with the hermitian part of ${\cal H}_{\rm int}$ in Eq.~(\ref{eq:H-eff}).
At the start of the holding time,
the scattering length has its final value $a$ and
there is a specified initial total number of atoms $N_0$,
an unknown fraction $f_D$ of which are bound atoms forming the dimer condensate.
We consider atoms trapped in a cylindrically 
symmetric harmonic potential:
$V(\bm{r}) = \frac12 m \omega_z^2 [z^2 + \zeta^2 (x^2 + y^2)]$.
If $f_D$ is sufficiently small, the effect of the dimer condensate 
on the atoms is negligible.
If the ramp to the final scattering length is slow enough,
the atom condensate remains adiabatically in its ground state.
We assume $N_A a$ is large enough that the kinetic energy 
of the atoms is small compared to their potential and interaction energies.
The atom condensate then has the familiar Thomas-Fermi 
density profile:
\begin{equation}
n_A(\bm{r}) = 
\frac{m}{4 \pi \hbar^2 a} \max \{ 0, \mu_A - V(\bm{r})\}.
\label{eq:TF-atomBEC}
\end{equation}
The chemical potential  is determined by the number 
$N_A = (1-f_D) N_0$ of unbound atoms: 
$\mu_A = \frac12 m \omega_z^2 a_z^2 [15 \zeta^2 N_A a/a_z]^{2/5}$,
where $a_z = (\hbar/m \omega_z)^{1/2}$.

We next consider the initial condition for $d(\bm{r},0)$.
Since the initial number of dimers $f_D N_0/2$ is small compared to 
$N_0$, their self-interaction energy is negligible but the mean-field
energy from the atom condensate can be important.
Assuming the dimer condensate produced by the ramp of the 
scattering length is in its ground state, it satisfies the 
Schroedinger equation
\begin{eqnarray}
\left[ - \frac{\hbar^2}{4m} \bm{\nabla}^2 + 2 V(\bm{r}) 
+ \frac{3 \pi \hbar^2\, {\rm Re} a_{AD}}{m} n_A(\bm{r}) \right] d(\bm{r})
\nonumber\\
= \mu_D d(\bm{r}).
\label{eq:SchEq-d}
\end{eqnarray}
The eigenvalue $\mu_D$ is the chemical potential of the dimer.
The normalization of the dimer condensate is determined
by the condition $\int\! d^3r \, d^* d = f_D N_0/2$.
The total potential energy of the dimer is the sum of the 
trapping potential $2 V(\bm{r})$ and the mean-field energy.
Its minimum is at the origin if ${\rm Re} a_{AD} < \frac83 a$
and near the edge of the atom condensate if ${\rm Re} a_{AD} > \frac83 a$.
In these two cases, the dimer condensate is an ellipsoid centered at the origin
and an ellipsoidal shell
near the edge of the atom condensate, respectively.
The boundaries in $a$ for the ellipsoidal shell region are
at $a_*$ and $2.86\, a_*$.

\begin{figure}[tb]
\centerline{\includegraphics*[width=8.5cm,angle=0,clip=true]{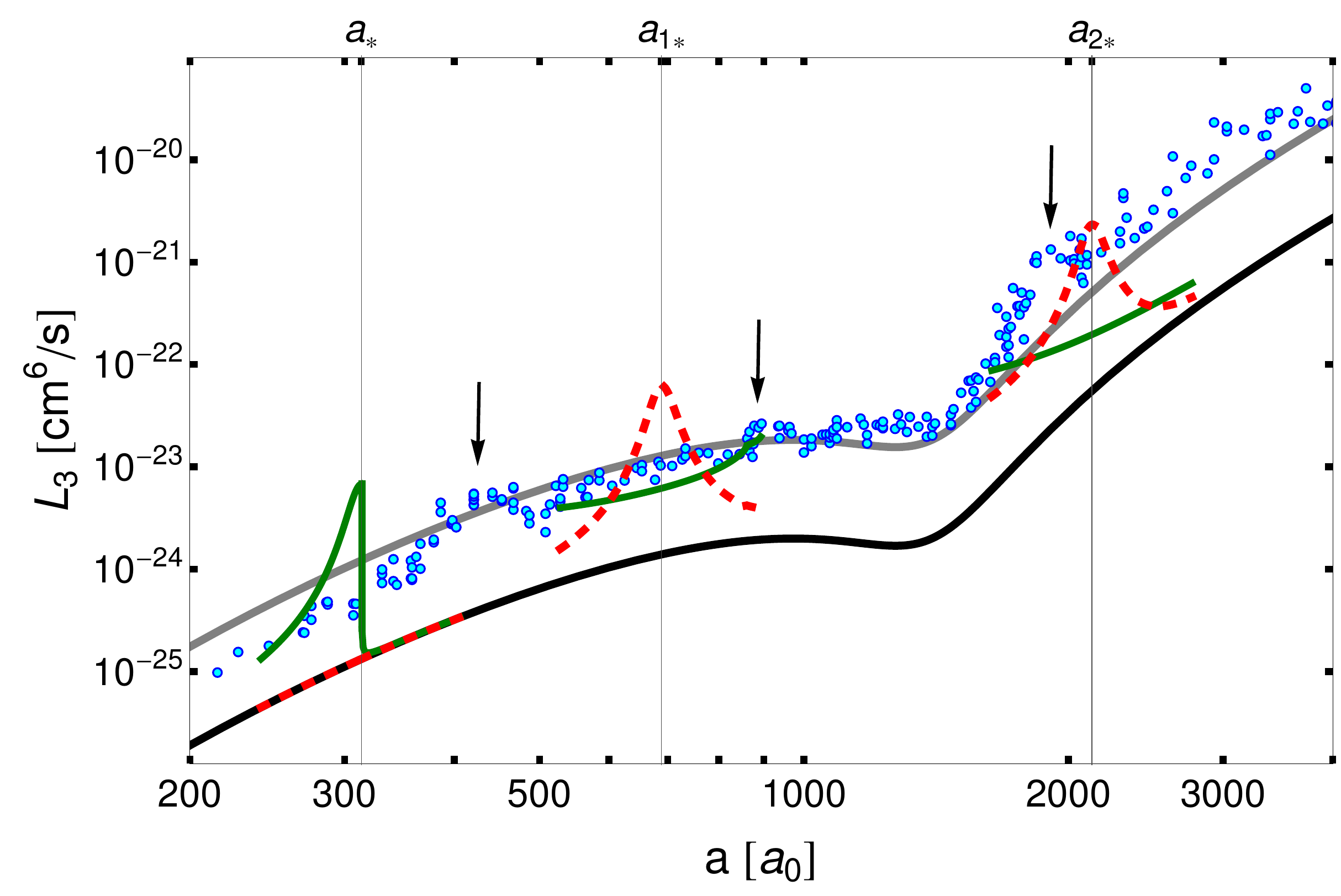}}
\vspace*{0.0cm}
\caption{ (Color online)
Three-atom recombination rate coefficient $L_3$
as a function of the scattering length $a$.
The data points were measured using a BEC of $^{7}$Li atoms \cite{Hulet:1301}. 
The arrows point to the narrow loss features identified in Ref.~\cite{Hulet:1301}. 
The lower and upper parallel curves are the universal result
for $L_3$ (with $a_+ = 1402~a_0$  and $\eta_* = 0.038$) and $9.2\, L_3$.
The three vertical lines are the universal predictions for $a_*$,
$a_{1*}$, and $a_{2*}$ using $a_+$ as input.
The additional atom-dimer and dimer-dimer contributions in $L_3^{\rm eff}$
are included in the thinner solid (green) and dashed (red) lines, respectively.
In the regions near $a_*$, $a_{1*}$, and $a_{2*}$, the dimer fractions  are
$f_D = 10^{-5}$, $3 \times 10^{-2}$, and $6 \times10^{-3}$.
}
\label{fig:Rice}
\end{figure}

The data on the loss rate of $^7$Li atoms in Ref.~\cite{Hulet:0911}
has recently been reanalyzed using a more accurate determination of the 
parameters of the Feshbach resonance \cite{Hulet:1301}.
Under the assumptions that the system is a pure BEC of atoms
and that the loss rate comes from 3-atom recombination only,
the rate coefficient $L_3$ is defined by 
$dN/dt = - (L_3/6) \int \!d^3r\, n_A^3$.
The data from Ref.~\cite{Hulet:1301} that was measured in a BEC 
of $^7$Li atoms is shown in Fig.~\ref{fig:Rice}.  
The initial number of atoms was $N_0 = 4 \times 10^5$.
Their fit to the universal prediction $L_3 = 3\alpha$ 
using an adjustable normalization factor is shown in Fig.~\ref{fig:Rice}.  
It determines the Efimov parameters $a_+ = 1402~a_0$ 
and $\eta_* = 0.038$ \cite{Hulet:0911} 
and requires a normalization factor of  approximately $9.2$.
The corresponding universal predictions for the resonances are
$a_* = 313~a_0$, $a_{1*} = 688~a_0$, and $a_{2*} = 2127~a_0$.
Near each of these three resonances, there is a narrow loss feature where 
the data are significantly higher than the fitted curve.
The positions of the loss features reported in Ref.~\cite{Hulet:1301}
are $426~a_0$, $919~a_0$, and $1902~a_0$.  
As shown in Fig.~\ref{fig:Rice}, 
they are not as widely spaced as the predicted resonances.

We now consider the effects of a small coexisting dimer condensate.
Our initial conditions are determined by $N_0 = 4 \times 10^5$ and 
an assumed dimer fraction $f_D$, which we expect to be small compared to 1 
and to depend strongly on $a$.
We take $n_A$ to be the Thomas-Fermi profile
in Eq.~(\ref{eq:TF-atomBEC}).  For $n_D = d^* d$, we use a
variational approximation for $d$ that reduces the 3-dimensional 
Schroedinger equation in Eq.~(\ref{eq:SchEq-d})
to a 1-dimensional equation for a single radial variable.
The subsequent time dependence of the total number of atoms
$N(t)$ can be obtained by solving Eqs.~(\ref{eq:dN/dt}).  
A quantitative comparison with the data would require 
comparing with the time dependence observed in the experiment.
Instead we compare the data for $L_3$ in Fig.~\ref{fig:Rice}
with an effective rate coefficient $L_3^{\rm eff}$ determined by the initial loss rate.
It is defined by $dN/dt = - (L_3^{\rm eff}/6)\int\!\! d^3r\, n_{A0}^3$,
where the integral is evaluated under the assumption that $f_D=0$:
$\int\!\! d^3r\, n_{A0}^3 = (N_0/168 \pi^2 a_z^4 a^2)[15 \zeta^2 N_0 a/a_z]^{4/5}$.
In Fig.~\ref{fig:Rice}, the universal prediction for $L_3$ from 3-atom recombination
falls an order of magnitude below the data.  This allows room for 
additional contributions 
from the atom-dimer and dimer-dimer terms in Eqs.~(\ref{eq:dN/dt}).  
We would like to determine whether narrow loss features can stand out 
above the smooth contributions for plausible values of $f_D$.

In Fig.\ref{fig:Rice}, the terms in $L_{3}^{\rm eff}$ are illustrated by two thin curves,
which correspond to 3-atom recombination plus atom-dimer losses 
and to 3-atom recombination plus dimer-dimer losses.  
We show these curves near $a_*$, $a_{1*}$, and $a_{2*}$ 
using different values of $f_D$ chosen to make the narrow loss feature visible: 
$f_D = 10^{-5}$, $3 \times 10^{-2}$, and $6 \times10^{-3}$, respectively.
The curve near $a_*$ that includes the atom-dimer terms in Eqs.~(\ref{eq:dN/dt})
has a narrow peak at the atom-dimer resonance.
The peak is  not symmetric, because the dimer condensate changes from
an ellipsoid centered at the origin for $a < a_*$
to an ellipsoidal shell near the edge of the atom condensate for 
$a > a_*$.  The curves near $a_{1*}$ and $a_{2*}$ that include the 
dimer-dimer term in Eqs.~(\ref{eq:dN/dt})
have narrow peaks at the two-dimer resonances.
The behavior near $a_{1*}$ and $a_{2*}$
is different, because  near $a_{1*}$ the dimer condensate 
is an ellipsoidal shell while near $a_{2*}$ it is an ellipsoid.
If $f_D$ is large enough near $a_{1*}$ and $a_{2*}$,
the peaks can stand out above 
the 3-atom recombination and atom-dimer contributions.  
If $f_D$ is too large near $a_+$, the atom-dimer contribution
can fill in the interference minimum from 3-atom recombination.
For $\eta_* = 0.038$, there is no longer a local minimum near $a_+$ 
if $f_D > 2.2 \times 10^{-4}$. 
The dimer fractions illustrated in Fig.~\ref{fig:Rice} are sufficient to make the 
atom-dimer and two-dimer resonances stand out in the initial loss rate.
Larger fractions would be required to make a significant 
difference in the integrated loss rates.
Nevertheless our results demonstrate that narrow loss features can be 
produced at the atom-dimer and dimer-dimer resonances 
with plausibly small values of the dimer fraction.

Our dimer condensate mechanism provides a plausible explanation for 
the narrow loss feature at an atom-dimer resonance that was observed 
in a BEC of $^{39}$K atoms \cite{Zaccanti:0904} 
and for the narrow loss features near an atom-dimer resonance 
and near two two-dimer resonances that were observed 
in a BEC of $^7$Li atoms \cite{Hulet:0911}.
It cannot explain the narrow loss features 
near atom-dimer resonances that have been observed in thermal gases 
of $^{39}$K atoms \cite{Zaccanti:0904} 
and $^7$Li atoms \cite{Hulet:0911,Khaykovich:1201}.
These loss features were observed at relatively small scattering lengths, 
so they could be associated with nonuniversal effects.
In a BEC, the dimer fraction $f_D$ would depend sensitively on the 
detailed form of the ramp that brings the scattering length to its final value $a$.
This sensitivity could be exploited to test our mechanism.
The fraction $f_D$ could
be amplified by pulsing the magnetic field very close to the Feshbach
resonance before measuring the loss rate, as in the 
experiments on atom-molecule coherence \cite{Weinman:0204}.
The loss rates at atom-dimer and dimer-dimer resonances would increase 
linearly and quadratically with the amplification factor, respectively.

In summary, we have proposed a new mechanism for narrow atom loss
features at atom-dimer resonances and at two-dimer resonances
in a BEC of ultracold atoms.  
The positions of these features are determined by universal 
few-body physics and thus provide additional tests of universality.
There could be similar loss features where universal 5-atom clusters 
cross the atom-dimer-dimer threshold.
The strengths of all these loss features are determined by the many-body 
physics of Bose-Einstein condensates and open a new window 
into the remarkable phenomenon of atom-molecule coherence.

\begin{acknowledgments}
We thank R.~Hulet for valuable discussions.
This research was supported in part by a joint grant from 
the Army Research Office 
and the Air Force Office of Scientific Research.
\end{acknowledgments}


\begin{thebibliography}{99}

\bibitem{Braaten:2004rn}
  E.~Braaten and H.-W.~Hammer,
  Phys.\ Rept.\  {\bf 428}, 259 (2006)
  [arXiv:cond/mat0410417].

\bibitem{Efimov70}
V.~Efimov,
Phys.\ Lett.\ {\bf 33B}, 563 (1970).

\bibitem{Efimov73}
V.~Efimov,
Nucl.\ Phys.\ A {\bf 210}, 157 (1973).

\bibitem{Platter:2004qn} 
  L.~Platter, H.W.~Hammer and U.-G.~Meissner,
  Phys.\ Rev.\ A {\bf 70}, 052101 (2004)
  [cond-mat/0404313].
  
\bibitem{Hammer:2006ct} 
  H.-W.~Hammer and L.~Platter,
  Eur.\ Phys.\ J.\ A {\bf 32}, 113 (2007)
  [nucl-th/0610105].
  
\bibitem{vSIG:0810} 
J.~von Stecher, J.P.~D'Incao, and C.H.~Greene,
Nature Physics {\bf 5}, 417 (2009)
[arXiv:0810.3876].

\bibitem{vonStecher:1106} 
J.~von Stecher,
Phys.\ Rev.\ Lett.\ {\bf 107}, 200402 (2011)
[arXiv:1106.2319].

\bibitem{Grimm:0512}
T.~Kraemer et al.,
Nature {\bf 440}, 315 (2006)
[arXiv:cond-mat/0512394].

\bibitem{Grimm:0903} 
F.~Ferlaino et al.,
Phys.\ Rev.\ Lett.\  {\bf 102}, 140401 (2009)
[arXiv:0903.1276].

\bibitem{Grimm:1205}
A.~Zenesini et al.,
arXiv:1205.1921.

\bibitem{Grimm:0807}
S.~Knoop et al.,
Nature Physics {\bf 5}, 227 (2009)
[arXiv:0807.3306].

\bibitem{Ferlaino:1108}
F.~Ferlaino et al.,
Few-Body Syst.\ {\bf 51}, 113 (2011)
[arXiv:1108.1909].

\bibitem{Zaccanti:0904}
M.~Zaccanti et al., 
Nature Physics {\bf 5}, 586 (2009) 
[arXiv:0904.4453].
    
\bibitem{Hulet:0911}
S.E.~Pollack, D.~Dries, and R.G.~Hulet,
Science {\bf 326}, 1683 (2009)
[arXiv:0911.0893].

\bibitem{Khaykovich:1201}
O.~Machtey et al.,
Phys.\ Rev.\ Lett.\ {\bf 108}, 210406 (2012)
[arXiv:1201.2396].
  
\bibitem{LSB:1205}
C.~Langmack, D.H.~Smith, and E.~Braaten,
Phys.\ Rev.\ A {\bf 86}, 022718 (2012) 
[arXiv:1205.2683].

\bibitem{LSB:1209}
C.~Langmack, D.H.~Smith, and E.~Braaten,
Phys.\ Rev.\ A {\bf 87}, 023620 (2013) 
[arXiv:1209.4912].

\bibitem{Weinman:0204}
E.A.~Donley, N.R.~Claussen, S.T.~Thompson, and C.E.~Wieman, 
Nature {\bf 417}, 529 (2002)
[arXiv:cond-mat/0204436].

\bibitem{KH:0204}
S.J.J.M.F.~Kokkelmans and M.J.~Holland, 
Phys.\ Rev.\ Lett.\ {\bf 89}, 180401 (2002)
[arXiv:cond-mat/0204504].

\bibitem{GKB:0209}
T.~K\"ohler, T.~Gasenzer, and K.~Burnett, 
Phys.\ Rev.\ A {\bf 67}, 013601 (2003)
[arXiv:cond-mat/0209100].

\bibitem{DS:0312}
R.A.~Duine and H.T.C.~Stoof,
Phys.\ Rep.\ {\bf 396}, 115 (2004)
[arXiv:cond-mat/0312254].

\bibitem{Deltuva:1107} 
  A.~Deltuva,
  Phys.\ Rev.\ A {\bf 84}, 022703 (2011)
  [arXiv:1107.3956].

\bibitem{Hulet:1301}
P.~Dyke, S.E.~Pollack, and R.G.~Hulet,
arXiv:0302.0281.

\end{thebibliography}
\end{document}